\documentclass[twocolumn,preprintnumbers,prl,aps,amssymb,superscriptaddress]{revtex4}
\usepackage{graphicx}
\usepackage{dcolumn}
\usepackage{bm}

\begin{document}

\title{Tunneling Spectroscopy on $c$-axis Y$_{1-x}$Ca$_x$Ba$_2$Cu$_3$O$_{7-\delta}$ Thin Films}

\author{J.~H.~Ngai}
\affiliation{Department of Physics, University of Toronto, Toronto,
Ontario, M5S1A7 Canada}
\author{W.~A.~Atkinson}
\affiliation{Department of Physics and Astronomy, Trent University, Peterborough, Ontario K9J 7B8 Canada}
\author{J.~Y.~T.~Wei}
\affiliation{Department of Physics, University of Toronto, Toronto,
Ontario, M5S1A7 Canada}

\date{\today}


\begin{abstract}

Scanning tunneling spectroscopy was performed on $c$-axis Y$_{1-x}$Ca$_x$Ba$_2$Cu$_3$O$_{7-\delta}$ thin films for \emph{x}= 0, 0.05, 0.15 and 0.20 at 4.2K. The measured spectra show main-gap, sub-gap and satellite features which scale similarly in energy \emph{versus} Ca-doping, suggesting that they are associated with a single pairing energy. The data is analyzed with a multiband tunneling model which attributes the sub-gap features to the chain band and the satellite and main-gap features to the plane band for $d_{x^2-y^2}+s$ pairing symmetry. These results suggest that the superconductivity in Y$_{1-x}$Ca$_x$Ba$_2$Cu$_3$O$_{7-\delta}$ involves multiple bands.

\end{abstract}

\maketitle


Common to all the copper-oxide based high-temperature superconducting compounds are the CuO$_2$ planes. The CuO$_2$ planes are believed to be responsible for the wealth of phenomena observed in these materials \cite{Orenstein_Millis}, particularly the formation of Cooper pairs with $d_{x^2-y^2}$ symmetry \cite{vanHarlingen_Tsuei}. YBa$_2$Cu$_3$O$_{7-\delta}$ (YBCO) is peculiar among the cuprates in that it also has quasi-one-dimensional, metallic CuO chains, which may contain finite superfluid density below the superconducting critical temperature (\emph{T$_c$}) as suggested by various bulk probes \cite{Basov_95,Hardy,Gagnon_Taillefer}. Since long-range order cannot be easily sustained in one-dimension, the apparent presence of superfluid density in the chains suggests that the plane and chain bands are electronically coupled. Such coupling would imply that the superconductivity in YBCO is essentially multiband in nature \cite{Mazin_Andersen,Tachiki,ODonovan_Carbotte,Atkinson_99,Morr_Balatsky,Edwards_deLozanne}, and could conceivably affect its pairing symmetry. In fact, recent pair-tunneling experiments \cite{Tsuei_d_plus_s,Hilgenkamp} have revealed a two-fold, $d_{x^2-y^2}+s$ pairing symmetry \cite{Sun_Dynes,Kouznetsov_Dynes} in optimally-doped YBCO, with the $d$-wave node lines rotated away from the chain axis. 

At present it is not yet clear from these pair-tunneling experiments whether the observed $d+s$ pairing symmetry is intrinsic to the planes \cite{Lu_Shen_ARPES}, or it is an effect of coupling between the plane and chain bands \cite{Mazin_Andersen,Tachiki,ODonovan_Carbotte,Atkinson_99,Morr_Balatsky}. Quasiparticle tunneling spectroscopy could help to elucidate this issue, by revealing multigap features in the excitation spectrum and by providing information about the pairing symmetry. For example, quasiparticle tunneling experiments on MgB$_2$ have revealed multiple \emph{s}-wave gaps arising from multiband coupling \cite{Iavarone_MgB2,Schmidt_MgB2}. In the case of YBCO, c-axis tunneling spectroscopy experiments have revealed multiple spectral features in addition to the predominant $d$-wave gap \cite{Valles_Dynes,Maggio_Aprile,Cucolo,Wei_PRL_98}. Since multigap spectra could arise from multiband coupling, it would be helpful to understand the origin of these additional spectral features. Furthermore, since both multiband coupling in general \cite{Binnig_STO} and pairing symmetry in the cuprates may vary with carrier doping \cite{Tajima_05}, a detailed study of the spectral evolution with doping could yield important insights on the pairing in YBCO. 

In this Letter, we present a scanning tunneling spectroscopy (STS) study of Ca-doped $c$-axis YBCO films at 4.2K. The measured spectra show main-gap, sub-gap and satellite features which scale similarly in energy as a function of Ca-doping, suggesting that they are associated with a single pairing energy. The data is analyzed with a generic tunneling model which indicates that the main-gap and sub-gap features can be associated with the plane and chain bands respectively, while the satellite feature can be attributed to $d+s$ splitting of the gap maximum in the plane band.  These results suggest that the superconductivity in YBCO involves both the plane and chain bands.

The Y$_{1-x}$Ca$_x$Ba$_2$Cu$_3$O$_{7-\delta}$ thin films measured were epitaxially grown by pulsed laser-ablated deposition (PLD) to $\sim$50 nm thickness on \{100\}-oriented SrTiO$_{3}$ substrates. The films were made from targets with nominal \emph{x} = 0, 0.05, 0.15, 0.20 Ca-doping concentrations. Two to three films were made for each doping concentration resulting in average measured $T_c$ values of 89K, 85K, 81K and 78K respectively. Cation substitution of Ca$^{2+}$ for Y$^{3+}$ in the unit cell is known to introduce additional holes into the CuO$_2$ planes, thus overdoping the films and lowering the $T_c$ \cite{Bottger}. The films were transferred from the PLD chamber to the STS apparatus with less than 10 minutes exposure to air. STS was performed at 4.2K with either chemically-etched or mechanically-sheared Pt-Ir tips, by measuring the tunneling current $I$ versus sample bias voltage $V$. The measured \emph{I}-\emph{V} spectra were numerically differentiated to obtain the conductance $dI/dV$ spectra.

\begin{figure}[t]
\includegraphics[width=8cm] {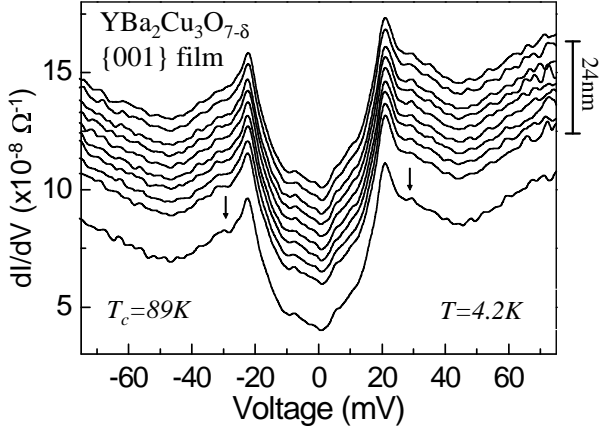}
\caption{\label{Optimalfigure}(color online) Representative tunneling spectra taken on an optimally-doped $c$-axis YBa$_2$Cu$_3$O$_{7-\delta}$ thin film at 4.2K. The spectra shown (staggered for clarity) were taken 3nm apart. Outer arrows indicate the satellite features.}
\end{figure}

Figure \ref{Optimalfigure} shows representative $dI/dV$ spectra taken at 4.2K on an optimally-doped ($x$= 0) $c$-axis film. The spectra, staggered for clarity, were taken every 3 nm over the film surface, demonstrating good spatial homogeneity. A robust main-gap feature is observed along with a weaker above-gap, satellite feature indicated by arrows in the bottom spectrum of Fig. \ref{Optimalfigure}. A weak sub-gap feature is also visible, becoming more pronounced with Ca-doping as will be discussed below.  All three spectral features have often been seen in previous $c$-axis tunneling experiments on YBCO single crystals and thin films, by both planar-junction and STS geometries \cite{Valles_Dynes,Maggio_Aprile,Wei_PRL_98,Cucolo,Ngai_PRB}. 

\begin{figure}[t]
\includegraphics[width=8cm] {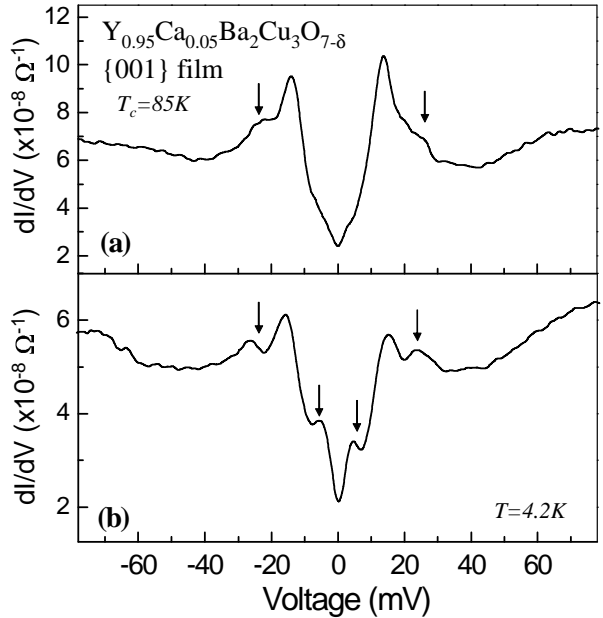}
\caption{\label{Ca5figure} (color online) Representative tunneling spectra taken on a $c$-axis Y$_{0.95}$Ca$_{0.05}$Ba$_2$Cu$_3$O$_{7-\delta}$ thin film at 4.2K.  (a) and (b) were taken on different spots on the same film, showing some spectral variation but the same generic features and similar main-gap sizes.  Inner and outer arrows indicate the sub-gap and satellite features respectively.}
\end{figure}
\begin{figure}[t]
\includegraphics[width=8cm]{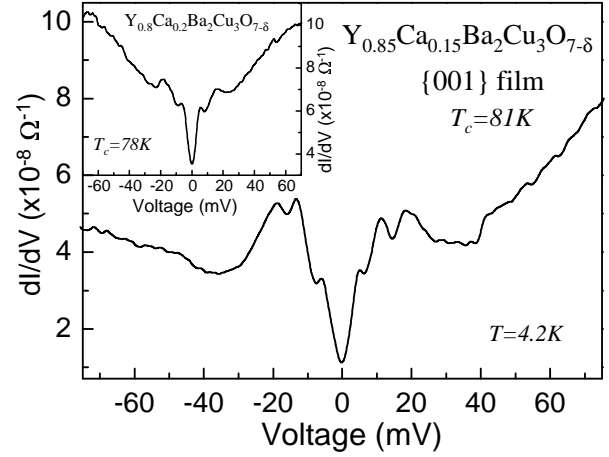}
\caption{\label{Ca15figure}(color online) Representative tunneling spectrum taken on a $c$-axis Y$_{0.85}$Ca$_{0.15}$Ba$_2$Cu$_3$O$_{7-\delta}$ thin film at 4.2K. (inset) Tunneling spectrum taken on a $c$-axis Y$_{0.80}$Ca$_{0.20}$Ba$_2$Cu$_3$O$_{7-\delta}$ thin film at 4.2K showing one of the larger main-gaps measured.}
\end{figure}

Figure \ref{Ca5figure} shows two types of spectra measured on different locations over a $x$= 0.05 Ca-doped film. Both spectral types exhibit the main-gap peaks, as well as the satellite features, while the spectrum in \ref{Ca5figure}(b) also exhibit pronounced sub-gap features, as indicated by the arrows inside the main-gap. We also note that the main-gap peak heights are more pronounced in Fig.\ref{Ca5figure}(a) than in Fig.\ref{Ca5figure}(b). In general, all the Ca-doped YBCO thin films show spatial variation in the STS spectra over distances beyond $\sim$20nm. This inhomogeneity was also observed in previous STS studies of similar Ca-doped YBCO thin films, indicating that this may be an effect of Ca substitution \cite{Deutscher_Review,Yeh_PRL}. Despite this spectral variation, the same generic features were consistently seen in our films up to $x$= 0.15. The main panel of figure \ref{Ca15figure} plots data for a \emph{x} = 0.15 film, again showing the main-gap, sub-gap and satellite features, indicating their prevalence in Ca-doped YBCO. 

\begin{figure}[t]
\centering
\includegraphics[width=8cm] {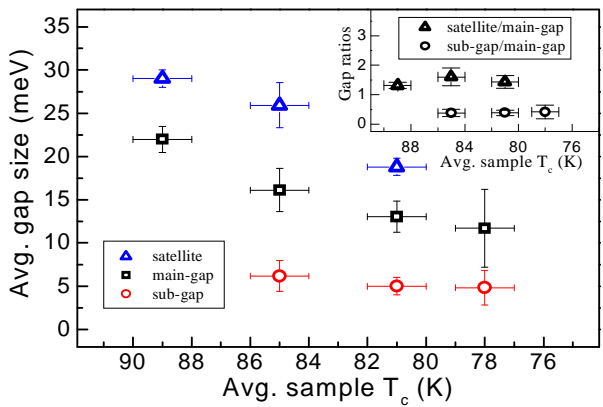}
\caption{\label{Ca20figure}(color online) Average gap sizes, corresponding to the satellite (triangle), main-gap (square), and sub-gap (circle) features, plotted as a function of the average sample $T_c$. Inset shows the average gap ratios, between satellite to main-gap (triangle) and sub-gap to main-gap (circle).}
\end{figure}

Figure \ref{Ca20figure} shows the evolution of these spectral features with Ca-doping. Assuming these features correspond to gap energies, a general trend of decreasing gap size versus increasing doping is observed. Quite remarkably, these spectral features show similar scaling in energy as a function of doping, as indicated by the roughly constant ratios of either satellite to main-gap, or sub-gap to main-gap in the inset of Fig.\ref{Ca20figure}. While there are several scenarios that could lead to multiple peaks in the tunneling spectrum \cite{Atkinson_99,Atkinson_unpub}, this apparent energy scaling suggests that there is a {\em single} pairing energy responsible for all three spectral features. This pairing energy presumably corresponds to the predominant pairing interaction in the CuO$_2$ planes.

In order to understand how multiple spectral features could arise from a single pairing energy in YBCO, we discuss a multiband scenario in which the CuO chain band is coupled to the CuO$_2$ plane band with an overall $d_{x^2-y^2}$+$s$ pairing symmetry. First, if one characterizes the order parameter in the plane by $\Delta_p(k) = \Delta_0[\cos(k_x)-\cos(k_y)]/2 + \Delta_s$ with the ratio $\Delta_s/\Delta_0 \approx 0.15$ \cite{Tsuei_d_plus_s,Hilgenkamp}, then the $d$-wave gap peak splits into main and satellite peaks \cite{ODonovan_Carbotte}. Second, in chain-plane coupling scenarios with the pairing interaction being predominantly in the plane, the gap function in the chain $\Delta_c(k)$ is proportional to $\Delta_p(k)$, creating a sub-gap energy scale \cite{Atkinson_99,ODonovan_Carbotte,Morr_Balatsky}. We illustrate these points with a generic tunneling model consisting of a two-dimensional plane band and a one-dimensional chain band without explicit account of their coupling. In this simplified model, the normal-state plane dispersion is
\begin{eqnarray}\label{plane}
\xi_p(k)& = & -2t_1[\cos(k_x)+\cos(k_y)+2t^{\prime}\cos(k_x)\cos(k_y) {} 
\nonumber\\
& & {}+ t^{\prime\prime}(\cos(2k_x)+\cos (2k_y))]-\mu_1                   
\end{eqnarray}
where $t^\prime=$ -0.1, $t^{\prime\prime}=$ -0.25 and $\mu_1 = \mu - \epsilon_1$, where $\mu$ is the chemical potential and $\epsilon_1$ a constant potential associated with the planes. We describe the normal-state chain dispersion by
\begin{equation}\label{chain}
\xi_c = -2t_2\cos(k_y) - \mu_2
\end{equation}
where $\mu_2 = \mu - \epsilon_2$, and where the potential difference $\epsilon_2-\epsilon_1$ determines the relative offset between the bottoms of the plane and chain bands. The parameters $t_1$, $t_2$ represent the nearest-neighbor overlap integrals in the plane and chain bands, respectively. We calculate the superconducting dispersion for each band via the BCS relation $E_{p,c}^2(k) = \xi_{p,c}^2(k) + \Delta_{p,c}^2(k)$, where for the chain gap function we take the simplest possible ansatz $\Delta_c(k) = \Delta_p(k)$, which allows us to illustrate our model. In reality, the induced gap in the chains will have a more complicated structure \cite{Atkinson_99}, however our model presented here is sufficient to capture many of the essential features predicted by a more thorough treatment, which calculates the coupling between the plane and chain layers explicitly \cite{Atkinson_unpub}. 

\begin{figure}[t]
\includegraphics[width=8cm] {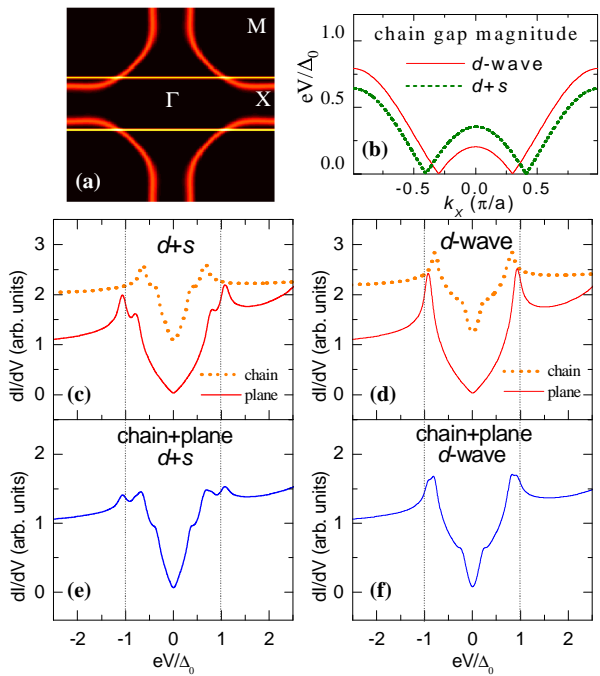}
\caption{\label{Analysisfigure}(color online) Two-band quasiparticle tunneling model used to describe the gap, sub-gap and satellite features. (a) Fermi surface used in our model with the 2D plane band shown in red along with the 1D chain band shown in orange. (b) Amplitudes of the \emph{d}-wave and $d+s$ gap functions projected on to the chain Fermi surface while moving along $k_x$. (c) Calculated $d+s$ spectra for the plane band (solid) and chain band (dashed) (d) Calculated \emph{d}-wave spectra for the plane band (solid) and chain band (dashed) (e) Combined plane and chain spectrum for $d+s$ scenario. (f) Combined plane and chain spectrum for pure \emph{d}-wave scenario.}
\end{figure}

We use the following parameter values $\{t_1, t_2, \mu_1, \mu_2\}$ = \{120, 300, -86, -350\} meV to calculate the Fermi surfaces of the plane (red) and chain (orange) bands shown in Fig.\ref{Analysisfigure}(a). Following the \emph{c}-axis quasiparticle tunneling formalism described in Ref. \cite{Wei_PRB}, we calculate the $dI/dV$ spectra for the plane (solid) and chain (dashed) bands using the $d+s$ gap function introduced above and show the results in Fig.\ref{Analysisfigure}(c). For comparison, we show the calculated spectra for pure $d$-wave gap function in Fig.\ref{Analysisfigure}(d). Due to the anisotropic nature of both the gap function and the band dispersions, the Fermi surface does not generally intersect with \emph{k}-space regions where the gap function is maximum. Figure \ref{Analysisfigure}(b) illustrates this scenario for the chain band, showing the $d$-wave and $d+s$ gap functions projected onto the chain Fermi surface along $k_x$. For the plane band, a similar scenario occurs. As a result of this \emph{band-structure enhanced} gap anisotropy, the $k$-space averaged gap edge measured by quasiparticle tunneling will tend to be smaller than $\Delta_0$. This tendency is shown in Fig.\ref{Analysisfigure}(c)-(f), where $\Delta_0$ is indicated by the dotted vertical lines. At present, our knowledge of the tunneling matrix elements between the tip and the individual plane and chain bands is incomplete \cite{Wei_PRB}, thus the relative weighting of each band in a given tunnel junction is unknown. To illustrate a multiband situation, we assume a 30\% plane to 70\% chain band weighting and plot a combined spectrum in Fig.\ref{Analysisfigure}(e) and (f) \cite{Iavarone_MgB2,Schmidt_MgB2,Cucolo}. For the parameters used above, the $d+s$ gap function produces three spectral features, while only two are produced for the \emph{d}-wave gap function. In comparing Fig.\ref{Analysisfigure}(e) with Fig.\ref{Ca15figure}, it is clear that the spectral features seen in our data can be attributed to a multiband scenario with $d+s$ pairing symmetry. 

On the premise that these spectral features arise from the plane and chain bands, a few remarks should be made on the origin of the $d+s$ pairing symmetry as well as doping evolution of both the $s$-wave component and the chain-plane coupling. First, O'Donovan and Carbotte \cite{ODonovan_Carbotte} have shown that the self-consistently solved order parameter pairing symmetry in a coupled plane and chain system is $d+s$. The predictions of their calculation are consistent with our experimental observations, and may in fact explain the origin of the $s$-wave component.   

Second, the 15\% $s$-wave component as reported in pair-tunneling experiments \cite{Tsuei_d_plus_s,Hilgenkamp} is within the error bars of our data up to \emph{x} = 0.15 Ca-doping. Beyond this doping level, the satellite feature is not discernible, as evidenced by the inset of Fig.\ref{Ca15figure}, showing one of the larger main-gaps observed on a \emph{x} = 0.20 film. Thus at present we cannot tell if the satellite feature has merged with the main-gap peak, or if the former has weakened to become indistinguishable from the spectral background. The scenario of the satellite and main-peak merging would suggest that the $s$-wave component has disappeared at the higher doping, leaving behind a purely $d$-wave pairing symmetry (Fig.\ref{Analysisfigure}(f)). Further studies using other experimental techniques would be necessary to elucidate this issue. 

Third, our data indicates that as YBCO is Ca-doped, the main-gap peaks are suppressed while the sub-gap features become more pronounced, suggesting that the chain-plane coupling could be changing with doping. This observation may be consistent with recent Raman scattering data showing an increase in the chain-plane coupling with Ca-doping \cite{Tajima_05}. Although our simple multiband tunneling model can explain the multiple spectral features we observed, a more rigorous theoretical treatment of the chain and plane coupling would be necessary to describe the peak-height evolution with doping \cite{Atkinson_99}.

Lastly, we note that tunneling studies of \{110\}-oriented Ca-doped YBCO thin films have observed a split zero-bias conductance peak (ZBCP) structure, which resembles the sub-gap features we observed on $c$-axis thin films \cite{Deutscher_Review}. However, the ZBCP splitting was observed to increase with Ca-doping \cite{Deutscher_Review}, in contrast to how the sub-gap feature evolves in our c-axis data. This difference clearly distinguishes the two junction orientations as probing different phenomena \cite{Wei_PRL_98}.

In summary, we have performed scanning tunneling spectroscopy measurements on $c$-axis Y$_{1-x}$Ca$_x$Ba$_2$Cu$_3$O$_{7-\delta}$ thin films at 4.2K, to study the evolution of the main-gap, sub-gap and satellite features with Ca-doping. These spectral features scale similarly in energy as a function of doping, suggesting that they are associated with a single pairing energy. Using a multiband tunneling model, our analysis indicates that the sub-gap and main-gap features can be attributed to the chain and plane bands respectively, while the satellite feature could arise from $d+s$ splitting of the gap maximum in the plane band. These results suggest that the superconductivity in Y$_{1-x}$Ca$_x$Ba$_2$Cu$_3$O$_{7-\delta}$ involves multiple bands.

\subsection{Acknowledgments}
\begin{acknowledgments}
This work was supported by grants from NSERC, CFI, OIT, MMO/EMK and the Canadian Institute for Advanced Research in the Quantum Materials Program. JHN would like to thank the WCS Foundation for funding. 
\end{acknowledgments}

\end{document}